\documentclass[fleqn]{annalen}
\usepackage{graphics}
\usepackage{latexsym}
\usepackage{amssymb}
\begin{document}
%%%%%%%%%%%%%%%%%%%%%%%%%%%%%%%%%%%%%%%%%%%%%%%%%%%%%%%%%%%%%%%%%%%%%%%%%%%%%%
%%%%%%%% the following newcommands will be completed by the publisher %%%%%%%%
%%%%%%%%%%%%%%%%%%%%%%%%%%%%%%%%%%%%%%%%%%%%%%%%%%%%%%%%%%%%%%%%%%%%%%%%%%%%%%
\newcommand{\volume}{8}              %sets current volume,
\newcommand{\xyear}{1999}            %sets year in header
\newcommand{\issue}{5}               %sets current issue,
\newcommand{\recdate}{29 July 1999}  %sets received date,
\newcommand{\revdate}{dd.mm.yyyy}    %sets revised date,     
\newcommand{\revnum}{0}              %number of revisions,
\newcommand{\accdate}{dd.mm.yyyy}    %sets accepted date,
\newcommand{\coeditor}{ue}           %sets (co)editor,
\newcommand{\firstpage}{0}         %first page number,  
\newcommand{\lastpage}{0}          %last page number,
\setcounter{page}{\firstpage}        %sets page counter to first page number 
%%%%%%%%%%%%%%%%%%%%%%%%%%%%%%%%%%%%%%%%%%%%%%%%%%%%%%%%%%%%%%%%%%%%%%%%%%%%%%
%%%%%%%%%%%%%%%%%%%%%%%%%%%%%%%%%%%%%%%%%%%%%%%%%%%%%%%%%%%%%%%%%%%%%%%%%%%%%%
%%%%%%%%%%%%%%%%%% please give up to three keywords here %%%%%%%%%%%%%%%%%%%%%
%%%%%%%%%%%%%%%%%%%%%%%%%%%%%%%%%%%%%%%%%%%%%%%%%%%%%%%%%%%%%%%%%%%%%%%%%%%%%%
\newcommand{\keywords}{localization in 2d systems, random flux model, chiral
symmetry, multifractality} 
%%%%%%%%%%%%%%%%%%%%%%%%%%%%%%%%%%%%%%%%%%%%%%%%%%%%%%%%%%%%%%%%%%%%%%%%%%%%%%
%%%%%%%%%%%%%%%% please give up to three PACS numbers here %%%%%%%%%%%%%%%%%%%
%%%%%%%%%%%%%%%%%%%%%%%%%%%%%%%%%%%%%%%%%%%%%%%%%%%%%%%%%%%%%%%%%%%%%%%%%%%%%%
\newcommand{\PACS}{71.30.+h, 73.40.Hm, 71.23.An, 72.15.Rn} 
%%%%%%%%%%%%%%%%%%%%%%%%%%%%%%%%%%%%%%%%%%%%%%%%%%%%%%%%%%%%%%%%%%%%%%%%%%%%%%
%% please enter (First) Author (et al.) and short version of the title here %%
%%%%%%%%%%%% must not exceed 80 characters in length together %%%%%%%%%%%%%%%%
%%%%%%%%%%%%%%%%%%%%%%%%%%%%%%%%%%%%%%%%%%%%%%%%%%%%%%%%%%%%%%%%%%%%%%%%%%%%%%
\newcommand{\shorttitle}{H. Potempa and L. Schweitzer, Localization in
correlated random magnetic fields} 
%% sets the header on oddpage
%%%%%%%%%%%%%%%%%%%%%%%%%%%%%%%%%%%%%%%%%%%%%%%%%%%%%%%%%%%%%%%%%%%%%%%%%%%%%%
%%%%%%%%%%%%%%%%%%%%%%%% here comes the title group %%%%%%%%%%%%%%%%%%%%%%%%%%
%%%%%%%%%%%%%%%%%%%%%%%%%%%%%%%%%%%%%%%%%%%%%%%%%%%%%%%%%%%%%%%%%%%%%%%%%%%%%%
\title{Localization of electrons in two-dimensional\\ spatially-correlated 
random magnetic fields}
%%%%%%%%%%%%%%%%%%%%%%%%%%%%%%%%%%%%%%%%%%%%%%%%%%%%%%%%%%%%%%%%%%%%%%%%%%%%%%
\author{H. Potempa and L. Schweitzer} 
%%%%%%%%%%%%%%%%%%%%%%%%%%%%%%%%%%%%%%%%%%%%%%%%%%%%%%%%%%%%%%%%%%%%%%%%%%%%%%
\newcommand{\address}
  {Physikalisch-Technische Bundesanstalt, Bundesallee 100, D-38116
  Braunschweig, Germany}
%%%%%%%%%%%%%%%%%%%%%%%%%%%%%%%%%%%%%%%%%%%%%%%%%%%%%%%%%%%%%%%%%%%%%%%%%%%%%%
\newcommand{\email}{\tt Ludwig.Schweitzer@ptb.de} 
\maketitle
%%%%%%%%%%%%%%%%%%%%%%%%%%%%%%%%%%%%%%%%%%%%%%%%%%%%%%%%%%%%%%%%%%%%%%%%%%%%%
\begin{abstract}
The localization properties of electrons moving in a plane
perpendicular to a spatially-correlated static magnetic field of
random amplitude and vanishing mean are investigated. 
We apply the method of level statistics to the eigenvalues and perform
a multifractal analysis for the eigenstates. 
From the size and disorder dependence of the variance of the nearest
neighbor energy spacing distribution, $P_{W,L}(s)$, a single branch scaling
curve is obtained. Contrary to a recent claim, we find no 
metal-insulator-transition in the presence of diagonal disorder.
Instead, as in the uncorrelated random magnetic field case, 
conventional unitary behavior (all states are localized)
is observed. 

The eigenstates at the band center, which in the absence of diagonal 
disorder are believed to belong to the chiral unitary symmetry class, 
are shown to exhibit a $f(\alpha)$-distribution for not too weak
random fields. The corresponding generalized multifractal 
dimensions are calculated and found to be different from the results 
known for a QHE-system.
\end{abstract}
%%%%%%%%%%%%%%%%%%%%%%%%%%%%%%%%%%%%%%%%%%%%%%%%%%%%%%%%%%%%%%%%%%%%%%%%%%%%%

\section{Introduction}
The localization properties of non-relativistic electrons moving in a
two-dimensional disordered system in the presence of static random magnetic
fields (RMF) with zero mean have been discussed controversially for more than
a decade. Investigations of Anderson localization of 2d electrons 
with random complex hopping matrix elements date back to the early 
eighties \cite{LF81}. 
The actual study of this model was particularly inspired by physical 
situations encountered in high-$T_{\textrm{c}}$ superconductors 
and for composite-fermions in the quantum Hall effect at
half filling factor.

Now, for lattice models of width $M$ and length $L$ with \textit{uncorrelated}
random magnetic fields the results gradually seem to converge 
to the notion that all states are localized
(see, e.g., \cite{Fur99} and papers quoted therein). 
In the absence of diagonal disorder, the localization length for
systems with $M$ odd is
believed to diverge at $E=0$ \cite{Fur99}, where the eigenstates are 
suggested to become critical \cite{MW96}. This special state at $E=0$
is a consequence of the chiral symmetry \cite{Zir96,AS99} 
which can be broken by diagonal disorder or, 
as reported in the case of quasi-1d systems, 
by application of periodic boundary conditions \cite{MBF99}.  
The random-flux model was shown to exhibit the same symmetry
\cite{Lea94,MW96} as the Gade-Wegner model \cite{Gad93}. 
The critical behavior is supposed to be similar to that of Dirac
fermions with random gauge fields \cite{Lea94,CMW96,HWK97}.

Recently, a metal-insulator transition has been asserted for the
lattice random-flux model incorporating \textit{spatially-correlated}
random magnetic fields and additional diagonal disorder \cite{SW99a}. 
This conclusion was based on a finite-size scaling study of the 
longitudinal conductance which was calculated numerically.  

The purpose of the present paper is twofold. First, we check the alleged 
metal-insulator transition by investigating the scaling behavior of the  
eigenvalue statistics. Second, the spatial fluctuations of the eigenstates
near energy $E=0$ are calculated and the scaling of the corresponding
moments are analyzed. In the presence of diagonal disorder, we find only
localized states for the spatially-correlated random flux model. 
Due to the huge localization length the wave functions near the band
center show multifractal behavior which, however, differs from the quantum 
Hall case.

\section{Model with correlated random flux}
A model of non-interacting electrons on a square lattice is considered
in a one-band tight-binding approximation. The spatially-correlated
random fluxes are generated as in Ref.~\cite{SW99a},
\begin{equation}
\phi(m)=4h_0/l_c^2 \sum_{n}R_n \exp(-|r_m-r_n|^2/l_c^2),
\end{equation}
where $h_0$ and $l_c$ are the strength and correlation length of the
random magnetic fields. The uncorrelated random numbers 
$R_n$ are evenly distributed between $(-1, +1)$ and the $r_m, r_n$
denote the spatial positions of the flux plaquettes considered.
The magnetic flux per plaquette, $\phi(m)$, enters the Hamiltonian via
the Peierls phase factors in the transfer terms from site $k$ to the
neighboring sites $k'$,
\begin{equation}
H=\sum_{k} \varepsilon_k\, c_k^\dagger c_k^{}+\sum_{\langle k\ne
k'\rangle} V \exp(i\theta_{kk'})\, c_k^\dagger c_{k'}^{}.
\end{equation}
The total flux through the plaquette labeled by $m$ is given by the sum
of the phases $\theta_{kk'}$ associated with its four links. 
$V=1$ is taken as the unit of energy and the lattice constant
$a=1$ as the unit of length. The diagonal disorder potentials
$\varepsilon_k$ are realized by a set of uncorrelated random numbers
taken from the interval $(-W, +W)$.
The eigenvalues have been calculated for square lattices of size 
$L=64$, 128, 192. In addition, some selected eigenstates 
were obtained for systems of size up to $L=305$.

\section{Results and discussion}
\begin{figure}
\centerline{\resizebox{8.2cm}{!}{\includegraphics{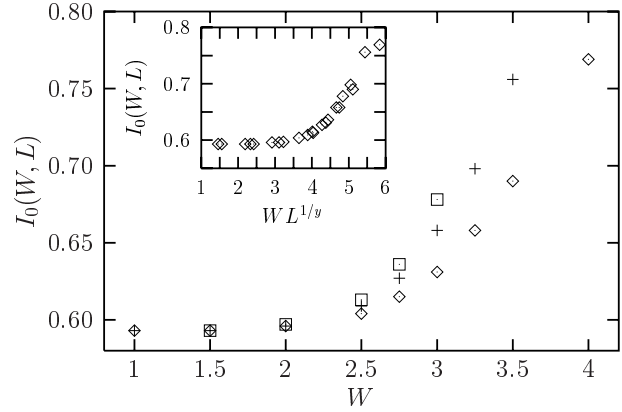}}}
\caption{Second moment of the level statistics, $I_0(W,L)$, versus
  disorder strength $W$. The system sizes are $L=64$ ($\Diamond$),
  $L=128$ ($+$), and $L=192$ ($\Box$). The inset shows the corresponding 
  single branch scaling curve.}  
\label{I0_WL}
\end{figure}

%\subsection{Level statistics}
The results of the level statistics are shown in Fig.~\ref{I0_WL} where the
second moment, $I_0(W,L)=1/2 \int_0^{\infty} P_{W,L}(s)\,ds$, 
of the level spacing distribution, $P_{W,L}(s)$, is plotted as a function
of the system size $L$ and disorder strength $W\!$. We take the same
parameters as in \cite{SW99a} so that the amplitude of the correlated
random fields $h_0=1.0$, the correlation length $l_c=5.0$, and the
Fermi energy $E_{\rm F}=-1.0$. The eigenvalues were calculated within
the interval $(E_{\rm F}-0.25, E_{\rm F}+0.25)$. Many realizations have been
considered such that the number of accumulated eigenvalues for each
pair ($L, W$) is $\approx 1\cdot 10^5$. 

We find that for $W>2$, $I_0(W,L)$ increases with increasing size
$L$. This behavior is expected for localized eigenstates which in the
limit $L\to \infty$ obey the Poisson statistics ($I_P=1.0$) of
uncorrelated energy levels. For $W<2$ no scale dependence is observed
for the  sizes investigated but $I_0(W,L)$ equals the random matrix
result $I_{\rm RMT}=0.59$. Therefore, the localization length
considerably exceeds the system size $L$.  If there were a critical
point below $W=2$ as suggested in \cite{SW99a}, $I_0(W_c,L)$ should be
scale independent at the critical disorder $W_c$ \textit{and} larger
than $I_{\rm RMT}$.  Hence, we have to conclude that for the suggested
parameters \cite{SW99a} there is no
localization-delocalization-transition in square systems.  The
corresponding  single branch scaling curve is shown in the inset of
Fig~\ref{I0_WL}. This result is similar to the situation found
previously for the uncorrelated magnetic field case \cite{BSK98}. 

The increase of the conductance for small $W$ in quasi-1d systems  
reported in \cite{SW99a} is presumably a finite size effect because
for narrow systems the increase in width $M$ may raise the conductance due
to the additional transport channels. In the limit of large $M$, however,
quantum interference will eventually localize the conductance. 

\begin{figure}
\centerline{%
\resizebox{7.cm}{!}{\includegraphics{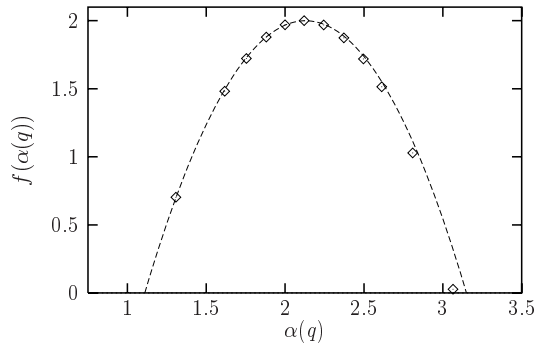}}}
\caption{ 
The $f(\alpha)$-distribution of an electronic eigenstate with energy 
close to the band center for a 2d system with correlated RMF. The 
parameters are $L=205$, $W=0$, $h_0=1$ and $l_c=5$.}  
\label{falfa}
\end{figure}

%\subsection{Multifractal eigenstates}
Turning now to the analysis of the eigenstates, we remind that the
spatial fluctuations of critical eigenstates
$\psi_{E,L}(r)$ exhibit multifractal properties, i.e., the scaling of 
the appropriate moments, 
$P_q(l,E,L)=\sum_i(\sum_{r\in \Omega_i(l)}|\psi_{E,L}(r)|^2)^q$, follow 
power-laws with an infinite set of unrelated exponents $\tau(q)=(q-1)D(q)$.
Alternatively, one considers the corresponding $f(\alpha)$-distribution
which for $|q|<1$ can often be approximated by 
$f(\alpha)=D(0)-(\alpha-\alpha_0)^2/[4(\alpha_0-D(0))]$, 
where  in 2d for our lattice system $D(0)=2$. Hence, $\alpha_0$
effectively characterizes the whole distribution. 

In Fig.~\ref{falfa} the $f(\alpha)$-distribution of an eigenstate near
$E=0$ of the correlated random magnetic flux model is shown. 
Similar results are obtained for 
% systems with an even number of lattice sites or for 
even $L$ or different $h_0$ and $l_c$ parameters as long as the
localization length considerably exceeds the system size. 
In most cases, a parabola can be fitted with 
$\alpha_0=2.14\pm0.03$. This number is quite distinct from the value 
$\alpha_0=2.29$ of a QHE-system \cite{HKS92}. From the scaling of the
second moment we find a fractal dimension $D(2)=1.87\pm0.03$ for the
correlated random magnetic field case which also differs from the QHE
value 1.6 \cite{HS94}.

Our values for \textit{spatially-correlated} RMF are similar to 
those reported for the \textit{uncorrelated} case. 
The authors of Ref.~\cite{KO95} determined $D(2)=1.8\pm0.06$ from the 
exponent $\delta$ of the temporal decay of the autocorrelation 
function using the relation $D(2)=2\delta$ \cite{HS94}, and $D(2)=1.79$ 
was extrapolated in Ref.~\cite{YG96} from a multifractal analysis
similar to ours. 
Multifractal wave functions have also been reported \cite{HWK97} 
for massless Dirac fermions near $E=0$ with a $f(\alpha)$-distribution
very close to the one shown here.

We find two exceptions to the multifractal behavior at $E\approx 0$ 
discussed above.
First, in the presence of diagonal disorder $W>1$
a tendency to strong localization can be observed which is fully
developed for $W>4$. Second, if the amplitudes of the spatially-correlated 
($l_c=5.0$) random magnetic fields are weak, $h_0\lesssim 0.2$, the 
eigenstates for odd system size $L\le 305$ and $W=0$ appear to be
extended ($\alpha_0 \approx 2.04$) which may be due to the small $L$
or an indication of the chiral unitary symmetry class.

\end{document}